# Measurement Noise Mitigation in a Quantum Computer Using Image Intensity Filters

Wladimir Silva, Dept. of Electrical and Computer Engineering, NC State University, Raleigh NC.

We propose a method to mitigate measurement errors in the distribution counts of a Quantum computer using image contrast filters. This work is similar to the method described by Gambetta and colleagues in [1]; however our technique does not use a linear system of equations, but an image contrast filter to mitigate the measurement noise. Furthermore, this method is demonstrated against the same set of experiments described in the matrix-free measurement mitigation (M3) library from Qiskit from which [1] is based upon. Our results show our method outperforming M3 by a wide margin in all experiments on IBM-Q. Furthermore, our method is platform agnostic; we demonstrate this by running some experiments on the IonQ cloud with similar results. Finally, we provide insights, documentation and detailed test and source code for further investigation.

## I. Introduction

Contrast filters are one of the simplest methods of image adjustment to improve its quality. Image contrast enhancement refers to accentuation or sharpening of image features so as to make it more useful for visualization or analysis. For a color image, the brightness of a pixel is the mean of its RGB values $= (r + g + b)/3$. This brightness can be increased by simply adding a delta value $\mu + \Delta\mu$ where the delta factor $\Delta\mu$ can be negative to darken the image or positive for lightning the image. A grayscale image, on the other hand, captures the intensity of light in pixels. Intensity values can be treated as floating point numbers ranging from 0-1. We propose a method to use this simple yet powerful technique to mitigate the noise of the histogram data of an experiment by first mapping its values to probabilities; then treating those probabilities as grayscale pixels. This allows for application of image processing analysis and display techniques with the ultimate goal of mitigating or eliminating the noise altogether.

## II. Method Description

Consider the first Greenberger–Horne–Zeilinger (GHZ) state describing a 3 qubit entangled state $|\psi\rangle = \frac{1}{\sqrt{2}}(|000\rangle + |111\rangle)$ as shown in see figure 1.

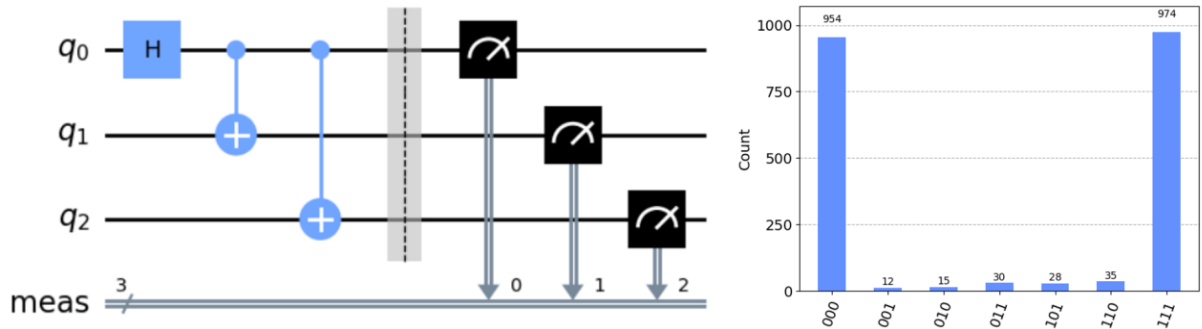

Figure 1. On the left, quantum circuit of a GHZ state, on the right experimental result from a run in IBM's 16 qubit noisy simulator FakeGuadalupe.

Take the measurement results of the experiment in figure 1 and construct a probability distribution by dividing each result count by the number of "shots" or measurements in the device; to obtain the distribution: $P(X) = [0.43847, 0.50390, 0.01611, 0.00585, 0.01220, 0.01513, 0.00732, 0.00097]$

We can map those probabilities to pixels in a Grayscale image. This allows for image manipulation and display. Next, apply a contrast filter to increase the separation between the darkest (low or noisy) and brightest areas of the image. The effect can be visualized in figure 2. Note that after the intensity rescaling step, the results need to be re-normalized resulting in a new distribution $P'(X) = [0.43454\ 0.50415\ 0.0.0.0.0.0.0.]$. Finally, map the new distribution into a new set of measurement results and compare against the original (right side of figure 2).

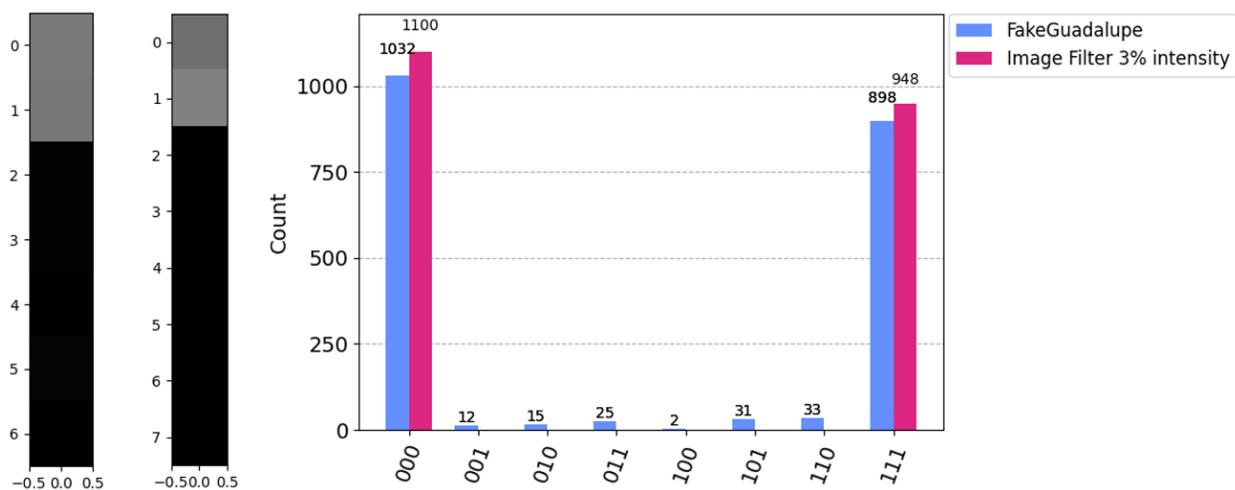

Figure 2. Visualization of the effect of brightness intensity rescaling and the final result for the GHZ state circuit in figure 1.

Note that the filter takes an optional intensity input range as argument to either stretch or shrink the intensity of the image. Thus given such an interval, values outside the interval are clipped to the interval edges. For example, if an interval of [0, 1] is specified, values smaller than 0 become

0, and values larger than 1 become 1. The results of the clipping effect are shown on the right side of figure2: the noise has vanished. This is a simple test yet encouraging enough to try this technique against the method described in [1] and implemented in the matrix-free measurement mitigation (M3) library by IBM research. Next, we run our method against the same set of tests described in the M3 documentation.

## III.  Experimental Results

We follow the scalable quantum measurement error mitigation tests described by the *mthree (M3)* library. M3 works by correcting a reduced subspace of input bit strings from an experiment measurement, resulting in a linear system of equations which is simpler to solve. For a small number of input strings the system of equations is solved using LU decomposition. For a large number of input strings the problem is solved in a matrix-free generalized minimal residual or bi-conjugate gradient stabilized method. The process is described in detail in [1], the set of original tests is available in [12], and our modified experiments are available in the source code section of this manuscript. The following sections showcase our data for probability distributions, sampling circuits, quantum resilience, VQE, and Ising models.

### 1. Probability Mitigation

We updated the experiment in [11] designed to correct readout errors and transform the outcomes to a true probability distribution. The experiment creates a circuit with 5 entangled qubits (center of figure 3), performs measurements in all of them, and runs in the 14 qubit noisy simulator FakeMelbourne collecting probability distributions for the raw device and M3. Our update includes image intensity mitigation at an input range of 3% (0.03, 0.97) (see figure 3).

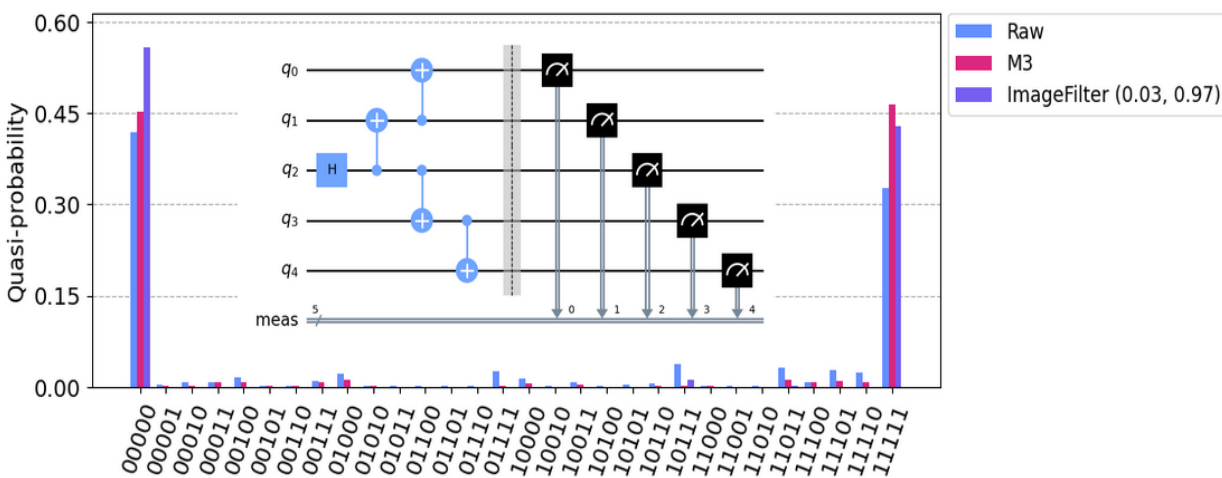

Figure 3. Experimental results for readout error correction in FakeMelbourne, modified from [11].

Our technique clearly outperforms M3 by reducing the noisy outcomes as shown in the probability histogram. Note that the good outcomes of the experiment are 00000, 11111 with 50% probability each. These are solid and encouraging results and the first step in a battery of tests to demonstrate the feasibility of this technique. Next we measure against M3's sampling experiments which yield even better results.

### 2. Sampling Mitigation with Bernstein-Vazirani (BV)

This is the mitigation test by sampling Bernstein-Vaziriani circuits of multiple qubit lengths as described in [2]. The test generates BV circuits for all-ones bit-strings of various lengths (see figure 4), and transpiles them against the 27 qubit noisy simulator FakeKolkata. This is an excellent test for mitigation techniques as the noise accrues when the number of Control-X gates increases.

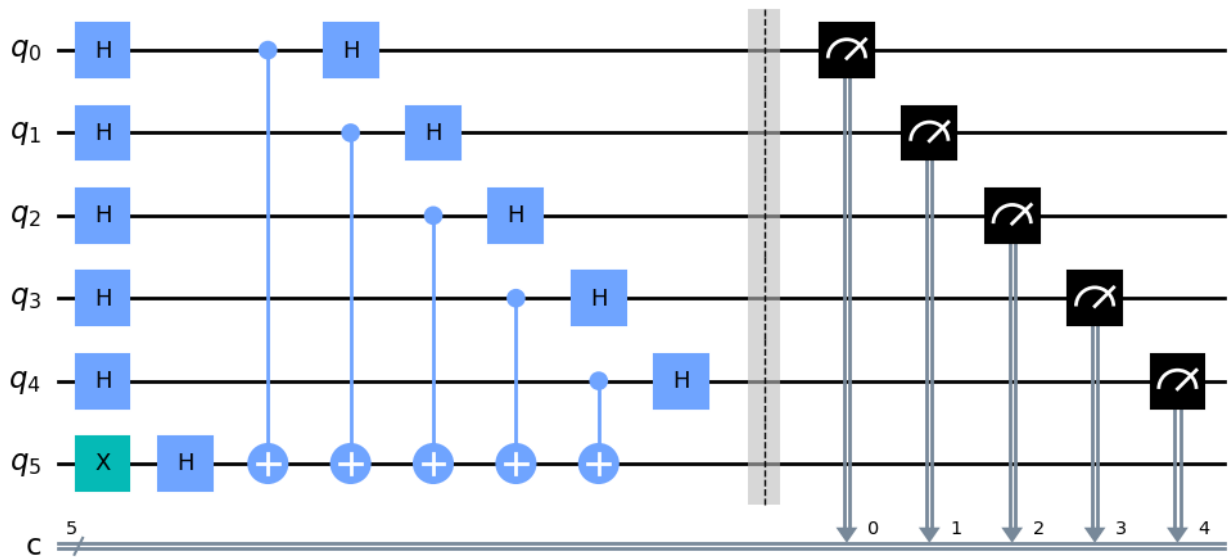

Figure 4. 5-qubit Bernstein-Vazirani circuit from the sampling mitigation test in [2].

We follow the test carefully, only altering the test code to insert the filter mitigation technique with intensity ranges of 1% [Input range (0.01, 0.99)] and 2% [Input range (0.02, 0.98)]. The success probability is displayed for the raw (unmitigated), M3 mitigated, and (1%, 2%) intensity levels (see figure 5).

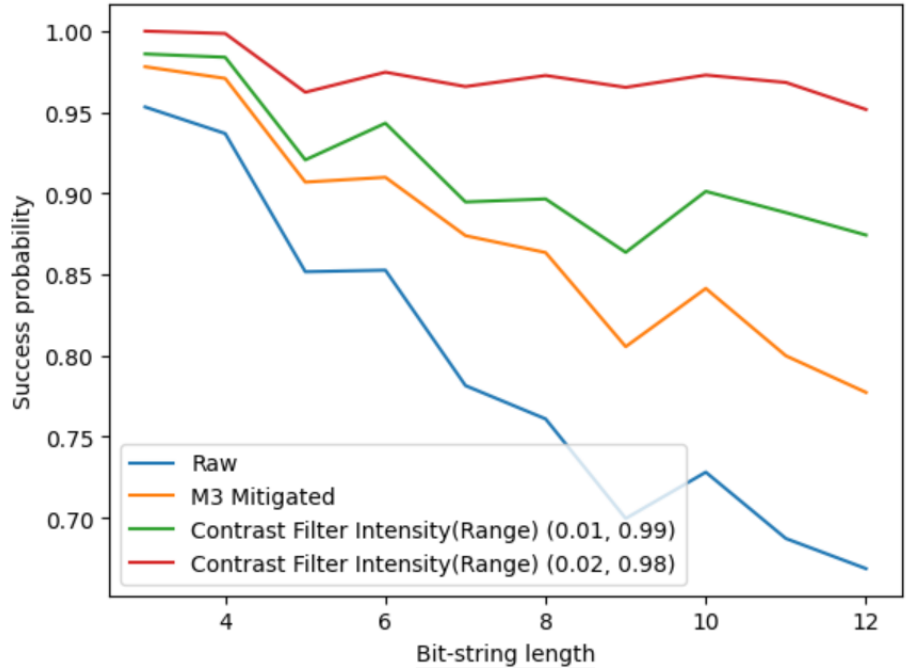

Figure 5. Mitigation experiment results for BV circuits of various lengths using M3.

The results show the intensity filter technique outperforming M3 by a solid margin for a 2% intensity level; even at 1% it produces better results. Furthermore, due to its simplicity, image manipulation filtering performs faster than the system of linear equations used by M3; this is described in more detail in section 4. The next test demonstrates correcting mid-circuit measurements using the dynamic version of the Bernstein-Vazirani algorithm.

### 3. Dynamic Bernstein-Vazirani

The dynamic Bernstein-Vazirani algorithm test uses Qiskit's sophisticated mid-circuit measurements and conditional resets to showcase M3's measurement error correction. This reduces the number of qubits significantly by allowing the same qubit to be reused after each partial measurement (top of figure 6). Note that the depth of the circuit will increase along with the bit strength which may affect the overall execution time. The experiment is run with the following parameters and the results are shown at the bottom of figure 6.

- Device: 27 qubit noisy simulator FakeKolkata.
- Measurement shots: 10000.
- Qubit strengths: 2-31.
- Mitigation: M3 vs Intensity filter at 1% input range of (0.01, 0.99).

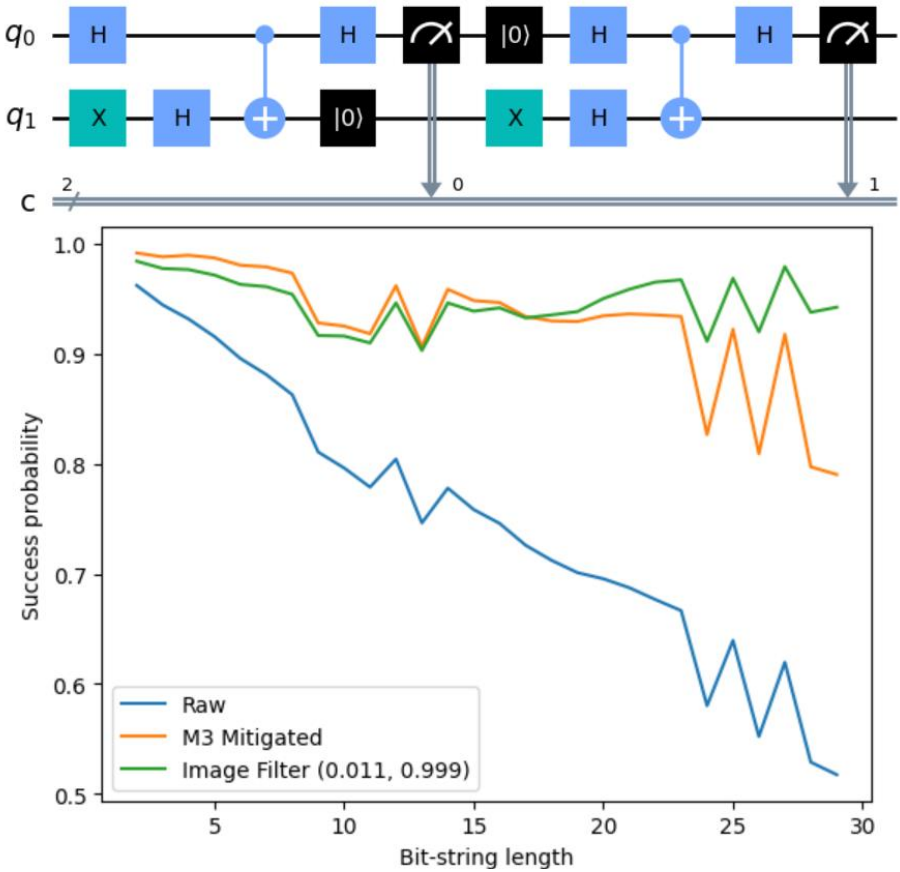

Figure 6. Dynamic Bernstein-Vazirani circuit for 2 qubits (top). Experimental results for dynamic BV circuits at multiple bit strengths (bottom).

After a modification of this test to include our technique, the result shows the intensity filter outperforming M3 as the number of qubits increases even at the low intensity level of 1%. Additionally, by increasing this level the noise can be virtually eliminated. In the next test, we go beyond the official M3 documentation to measure against an experiment that is part of the resilience mechanism implemented by IBM at the quantum level. Our technique shows promising results against quantum resilience.

### 4. Quantum Resilience on Trotterization

This experiment showcases the error suppression and mitigation at the quantum level built into the Qiskit Runtime [4]. This mechanism is tested against a set of Trotterization circuits which segment an exponentiated matrix into a series of smaller exponentiated matrices with minimal error. This method reduces the complexity by providing an estimate of the solution to a time-dependent Hamiltonian. For a detailed description of this technique see [5]. Qiskit runtime implements three levels of Quantum resilience:

1. Twirled readout error extinction (T-Rex): It is a general and effective technique to reduce measurement noise by attaching extra measurement and calibration circuits for the estimation of error mitigated averages [6].
2. Zero Noise Extrapolation (ZNE): It works by first amplifying the noise in the circuit of the desired quantum state, obtaining measurements for several different levels of noise, and using those measurements to infer the noiseless result [7].
3. Probabilistic error cancellation (PEC): It samples for a collection of circuits that mimic a noise inverting channel to cancel out the noise in the desired computation similar to the way noise canceling headphones work [7].

We altered the experiment to include the intensity filtering technique with the results shown in figure 8. Our technique outperforms T-Rex and ZNE by a good margin, and arguably does better than PEC (bottom left of figure 7). It should be noted that, although PEC performs the best among the quantum resilience techniques, it has a sampling overhead time complexity in the exponential level to the number of gates. As a matter of fact, by the time of this writing, some of our experiments using PEC resilience with circuits with more than 10 Control-X gates, crashed after multiple hours.

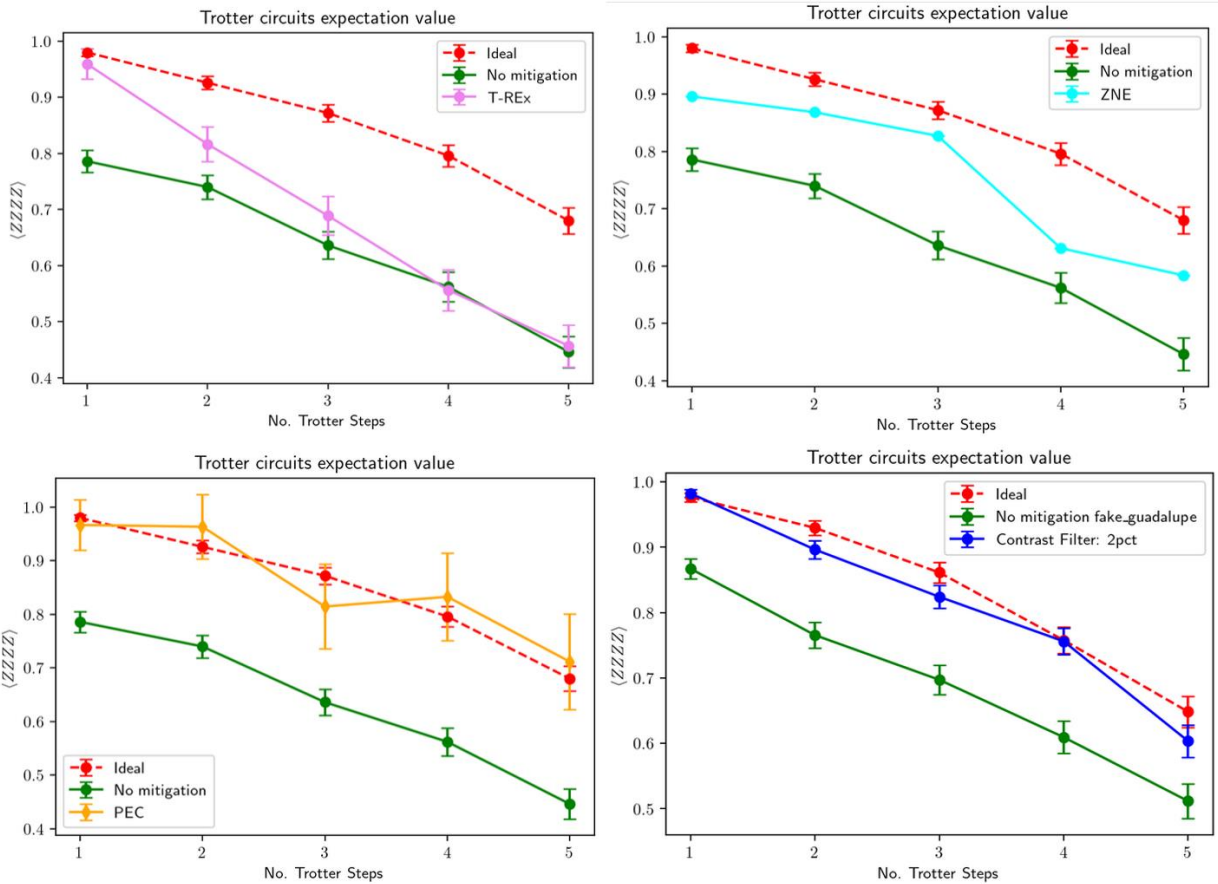

Figure 7. Results for the quantum resilience experiment using Trotterization circuits described in [5]. From the top left T-Rex, ZNE, PEC and image intensity filtering at 2% input range. Image filtering shows the closest to the ideal Trotterization result (bottom right).

## 5. Basic Variational Quantum EigenSolver (VQE)

This test is described in [8]. It takes the sample Hamiltonian $H = 0.3979YZ - 0.3979ZI - 0.01128ZZ + 0.1809XX$ and calculates the expected energy value using the following VQE state preparation:

- Ansatz: *TwoLocal* from Qiskit library with full entanglement (each qubit is entangled with all the others).
- Measurement circuits: Created by moving the observables in the Hamiltonian (YZ, ZI, ZZ, XX) into the computational basis by appending an Hadamard Gate (H) to each Pauli X gate, and a S-dagger and H gates for each Y gate in the observable with Z and I gates remaining unchanged as shown in figure 8.
- Initial parameter angles: [1.22253725, 0.39053752, 0.21462153, 5.48308027, 2.06984514, 3.65227416, 4.01911194, 0.35749589]

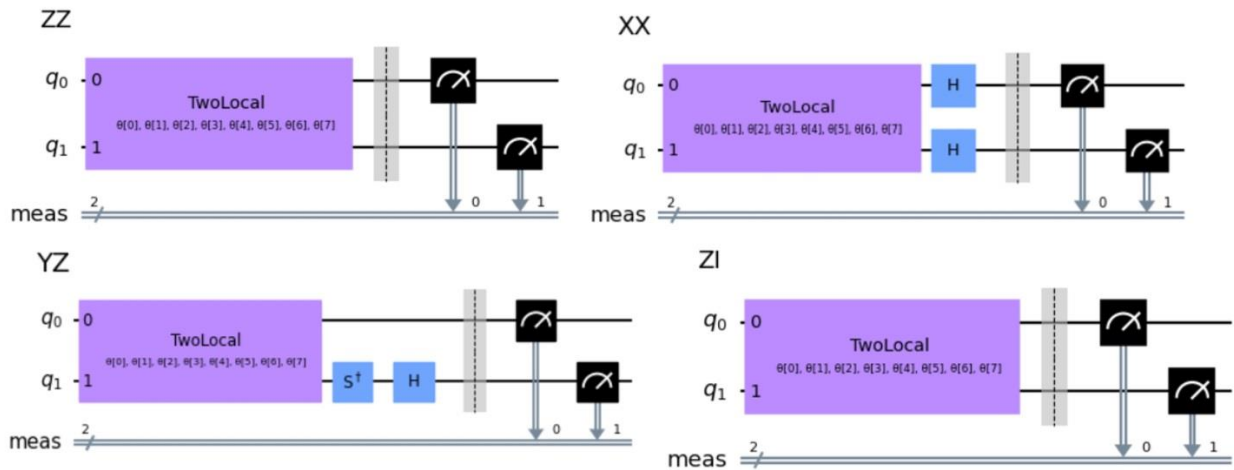

Figure 8. Measurement circuits for the VQE experiment in [8].

The image intensity filter method performs better than M3, yielding a more accurate estimate of the final energy of the Hamiltonian:

M3 Mitigation Results

- Energy: -0.4248273342124237
- Final angles: [1.25560585 0.38654154 0.20846952 5.48743517 2.03181833 3.68814594 3.91540956 0.36953303]

Image Intensity Filter Results at 1% input range (0.01, 0.99)

- Energy: -0.43324214733405414
- Final angles: [1.23757107 0.4211179  0.08658293 5.49275181 2.05102127 3.67389758 4.03722484 0.3746737]

The optimal solution is defined in [8]: -0.44841884382998787.

## 6. Ising Models using VQE

For the final experimental test, we chose the state preparation for the Kagome lattice using the VQE algorithm. This experiment was part of the IBM's Open Science Prize 2022. The challenge amounts to preparing the frustrated ground state of a Heisenberg spin-1/2 model on a Kagome lattice using the VQE algorithm. A basic notebook was provided by IBM for participants to do their work. We used this code to test the behavior of our method with encouraging results. The state as chosen for this experiment consists of:

- Hamiltonian: The antiferromagnetic Heisenberg model arranged on a Kagome lattice

$$H = \sum_{i,j} X_i X_j + Y_i Y_j + Z_i Z_j$$

- Ansatz: EfficientSU2 - a circuit made of layers of single qubit operations spanned by Unitary-Set-Dimension2 SU(2) and Controlled-X entanglements from Qiskit library. This is a heuristic pattern that can be used to prepare trial wave functions for quantum algorithms or classification circuits for machine learning. The default entanglement is set to *reverse_linear* as it provides fewer entangling gates than the rest.
- Backend: The 16-qubit noisy simulator FakeGuadalupe which closely mirrors IBM's hardware device ibmq_guadalupe.
- Optimizer: Nakanishi-Fujii-Todo algorithm [10] from Qiskit library. This is a technique that divides the parameterized quantum circuits into sub-problems using only a subset of the parameters; this transforms the cost function into a sine curve with period $2\pi$. By repeating this procedure, the parameterized quantum circuits are optimized and the cost function is minimized. For this test, the maximum number of iterations to perform defaults to 100.
- Shots or number of measurements in the experiment: 2048.
- The expected ground state: -18.

Because code modifications for this experiment require changes to the Qiskit runtime service, we were unable to run in a real device; however with a modified version of Qiskit's *BackendEstimator* class we injected our mitigation logic in the code. The results are presented in figure 10.

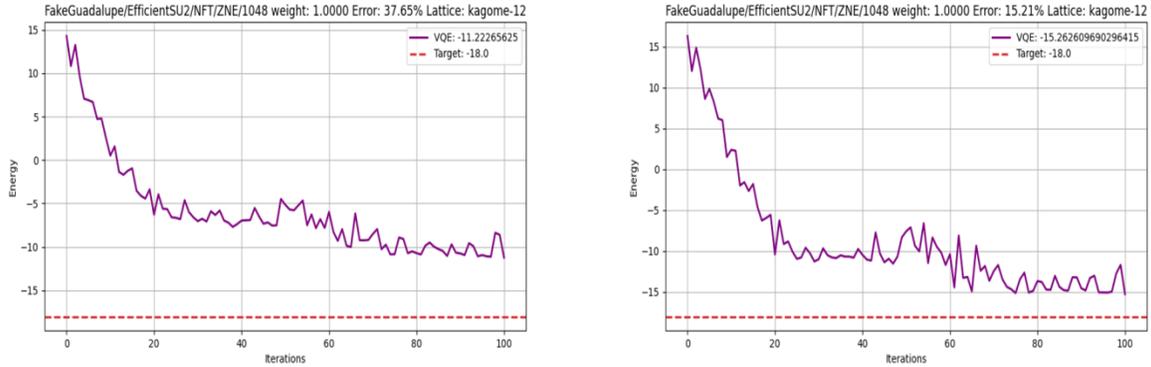

Figure 9. Experimental results for the ground state preparation of the Kagome lattice; the metric of performance in the challenge judging criteria if given by the relative difference between the expected ground state (-18.0) and the VQE result.

The left side of figure 9, shows the unmitigated ground state with a relative error of 37.6%, on the right side the intensity filter mitigated at 2%% (0.002-0.998) showing a slim improvement at 15.2% relative error. We expected better performance nevertheless a noise reduction of 15 percentage points is remarkable.

## IV.   Performance Metrics

We gathered the execution times of both methods to gain an insight on their performance. The tests were run on an Intel Core i5 CPU @ 2.60GHz with 8GB of memory and a 64-bit Windows OS. The results are shown in table 1.

Table 1. Execution times in *milliseconds* for some of the experiments from section 3. Note that experiments 3.4 and 3.6 using the Qiskit remote runtime service require server side changes and cannot be easily integrated.

| Experiment | M3 Calibration | M3 Correction | Image Filter Correction | Qubit range |
|---|---|---|---|---|
| 3.1 Probabilities | 452 | 10 | 0 | 5-qubit entanglement |
| 3.2 Sampling | 670 | 0 | 0 | 3-7 |
| 3.3 Dynamic BV | 77174 | 94 | 10 | 2-15 |
| 3.5 Basic VQE | 190 | 12572 | 12177 | N/A |

The image filter method shows better overall performance. Note that M3 has a calibration step that controls miscellaneous execution parameters. Our data shows that this calibration step degrades performance significantly as the number of qubits increases.

## V. Platform Independence

Platform independence is an important aspect for any effective error mitigation system because it enhances portability and flexibility. To the best of our knowledge, the M3 mitigator object targets a specific IBM backend, by first computing the calibration matrices for the given qubits and given number of shots. It then applies mitigation to a given dictionary of raw counts over the specified qubits. Our method, on the other hand, works by treating the measured probabilities as image pixels, and applying an intensity contrast filter to reduce the noise. Therefore, our system should work on any quantum platform. To illustrate this, we run some of the experiments in section 3 on the IonQ cloud with the results shown in figure 10.

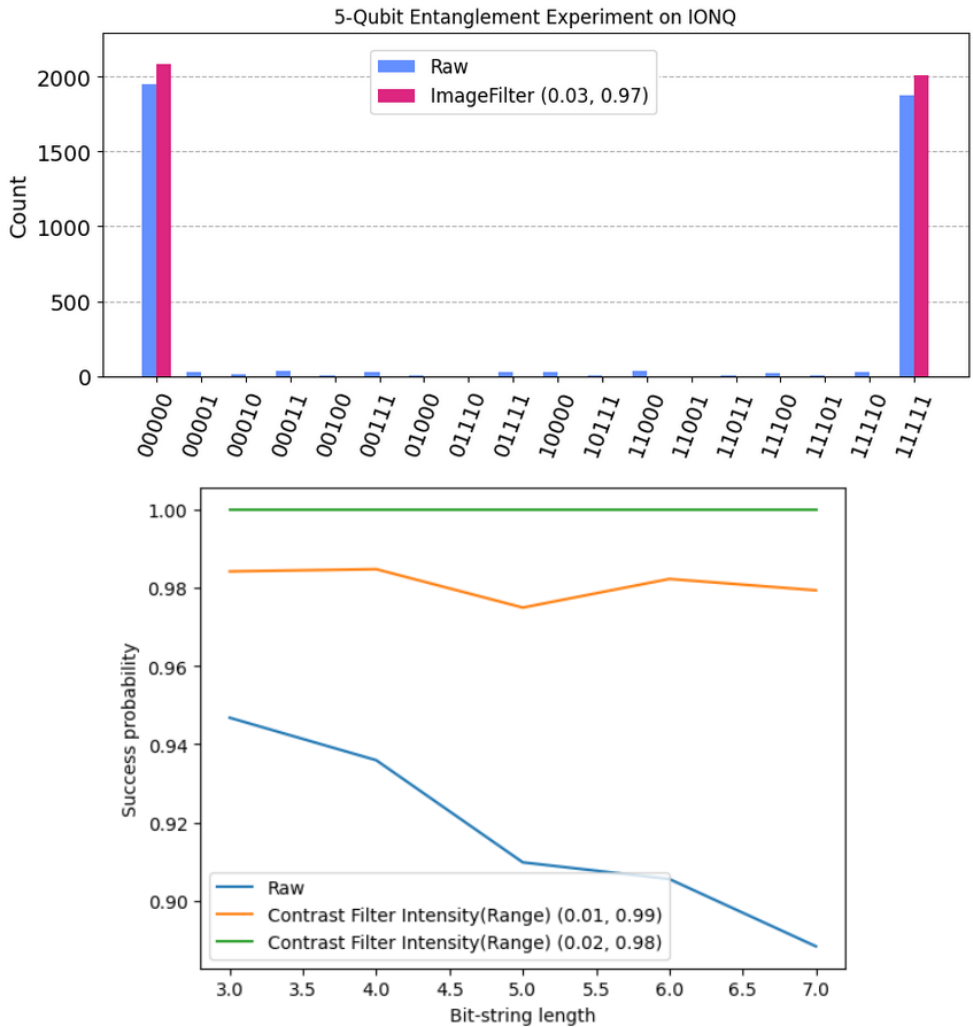

Figure 10. Results for experiments 3.1 (top) and 3.2 (bottom) on the IonQ cloud simulator with noise model *aria-1* which mimics IonQ's 25 qubit Aria-1 ion-trap quantum computer. The data shows results for the noisy (raw) probabilities vs the image filter at 3% intensity (top) and 1%, 2% (bottom).

## VI. Conclusion

We developed a method to mitigate measurement errors in the distribution counts of a Quantum computer using image contrast filters at various input ranges. This technique was extensively tested against IBM's sophisticated mitigation library M3 with remarkable results. Our method outperformed M3 by wide margins in all tests. We also tested against IBM's quantum resilience (T-Rex, ZNE and PEC) with equally outstanding outcomes. In this age of noisy quantum computers, error mitigation has become a prime area of research. It is critical to develop an efficient and fast method to eliminate noise from measurement outcomes. At the quantum level, there are many resilience techniques, error correction codes, and other methods; however, these approaches require an excessive number of ancillary qubits and furthermore contribute to the overall noise themselves. We believe, at this stage, noise can be managed at the classical level with a fast and efficient technique, and our method shows solid results in that direction. It outperforms the top implementation by IBM. The results speak for themselves.

## VII. Source Code

Detailed implementations for each experiment is available for download from GitHub at https://github.com/Shark-y/quantum_mitigation

## VIII. References

[1] "Scalable Mitigation of Measurement Errors on Quantum Computers", Paul D. Nation, Hwajung Kang, Neereja Sundaresan, and Jay M. Gambetta, PRX Quantum 2, 040326 (2021).

[2] Mitigating sampling problems, available online at
https://qiskit.org/ecosystem/mthree/sampling.html#

[3] Dynamic Bernstein–Vazirani available online at
https://qiskit.org/ecosystem/mthree/tutorials/04_dynamic_bv.html

[4] Error suppression and error mitigation with Qiskit Runtime, available online at
https://qiskit.org/ecosystem/ibm-runtime/tutorials/Error-Suppression-and-Error-Mitigation.html

[5] Grant Kluber. Trotterization in QM Theory.
https://web.ma.utexas.edu/users/drp/files/Fall2020Projects/DRP_Final_Project_F2020%20-%20Grant%20E%20Kluber.pdf

[6] Model-free readout-error mitigation for quantum expectation values. T-Rex
arXiv:2012.09738

[7] Error mitigation for short-depth quantum circuits, ZNE, PEC: arXiv:1612.02058

[8] Basic mitigated VQE, available online at
https://qiskit.org/ecosystem/mthree/tutorials/10_basic_vqe.html

[9] IBM Quantum Awards: Open Science Prize 2022. https://github.com/qiskit-community/open-science-prize-2022

[10] Sequential minimal optimization for quantum-classical hybrid algorithms. https://arxiv.org/abs/1903.12166

[11] Correcting probabilities available at https://qiskit.org/ecosystem/mthree/tutorials/02_correcting_probs.html

[12] mthree (2.5.1) user guide available on-line at https://qiskit.org/ecosystem/mthree/index.html